\documentclass[twocolumn]{aastex61}

\shorttitle{Oscillations in the first AO of Cycle 25}
\shortauthors{A.~Chelpanov and N.~Kobanov}
\begin{document}
\title{Multilevel Observations of the Oscillations in the First Active Region of the New Cycle}
\correspondingauthor{A.~\surname{Chelpanov}}
\email{chelpanov@iszf.irk.ru}
\author{Andrei~\surname{Chelpanov}}
\affil{Institute of Solar-Terrestrial Physics
                     of Siberian Branch of Russian Academy of Sciences, Irkutsk, Russia}
\author{Nikolai~\surname{Kobanov}}
\affiliation{Institute of Solar-Terrestrial Physics
                     of Siberian Branch of Russian Academy of Sciences, Irkutsk, Russia}

\begin{abstract}

For the first time, a multi-wave research of oscillation dynamics in a solar facula from its birth to decay was carried out.
 We performed spectral observations of active region NOAA\,12744 at \textit{Horizontal Solar Telescope} of the Sayan Solar Observatory in the H$\alpha$, He\,\textsc{i} 10830\,\AA, and Si\,\textsc{i} 10827\,\AA\ lines.
 We used \textit{Solar Dynamics Observatory} (SDO) line-of-sight magnetic field data and the 1600\,\AA, 304\,\AA, and 171\,\AA\ UV channels.
 At the early stages of the facula evolution, we observed low-frequency (1--2\,mHz) oscillations concentrate in the central part of the facula.
 In the lower solar atmosphere, this is registered in the intensity, line-of-sight velocity, and magnetic field signals.
 These frequencies were also observed in the transition region and corona (304\,\AA\ and 171\,\AA\ channels).
 At the maximal development phase of the facula evolution, the low frequency oscillations closely reproduce the coronal loop structures forming above the active region.
 At the decay phase, the spatial distributions of the observed frequencies resemble those found in and above the undisturbed chromosphere network.
 Our results indicate a direct relation of the low frequency oscillations observed in the lower solar atmosphere with the oscillations in the coronal loops, which is probably implemented through the loop footpoints.

\end{abstract}

\section{Introduction} \label{sec:intro}

Faculae are the most frequent expression of the solar activity.
They manifest themselves as large bright areas that harbor increased magnetic field flux.
They are always observed in active regions (AR), both containing sunspots and those not related to sunspots.
Faculae precede the emergence of sunspots and remain during the whole lifespan of ARs up to the sunspot decay and even after it.
%One also observes a large number of faculae that are not related to sunspot active regions.
Due to their surface area and abundance, faculae affect the energy balance between the solar atmosphere layers, and their influence includes that through the wave mechanisms.
Studying oscillation characteristics in faculae is a relevant problem of solar physics.
To date, numerous studies have addressed this problem \citep{1965ApJ...141.1131O, 1971SoPh...19..338S, 2008ApJ...676L..85K, 2011SoPh..268..329K, 2017A&A...598L...2K}.
Note that studying the spatial distributions of oscillations in faculae is more difficult than in sunspots due to a more complicated magnetic field topology: faculae often contain several separated magnetic knots of different polarities.
The evidence on the strength and direction of the magnetic field is highly diverse.
\citet{2010A&A...524A...3N} analysed polarimetric observations of a facula with a high spatial resolution and found a number of magnetic features: micropores, bright points, ribbons, flowers, and strings.
\citet{2019MNRAS.482.5290S} identified three main magnetic structures: small-scale flux tubes, knots, and pores.
\citet{1992ApJ...391..832R, 1997ApJ...474..810M, 2010SoPh..262...35G} claim that vertical magnetic fields dominate faculae, while \citet{2008A&A...481L..25I} found that small horizontal magnetic structures prevail in them.
\citet{2010A&A...509A..92B} based on the polar facula observations in the Fe\,\textsc{i} 1.5\,$\mu$m line found two magnetic structure types: one with a strength of 900--1500\,G and the other of strengths lower than 900\,G.
While the structures of the first type are vertical, the second type shows arbitrary angles.
The situation becomes even more complicated when one considers the changes in the characteristics following the evolution of a facula.
The first indications of a new solar activity cycle can be observed in the emerging faculae.

While the processes accompanying the evolution of AR have been discussed quite thoroughly in the contemporary publications \citep[][and references therein]{2015LRSP...12....1V}, studies of oscillation spectra dynamics during the facular AR evolution are almost non-existent. With this study we are trying to fill this gap.

This work is dedicated to multiwave studying of oscillations in a solar facula registered as NOAA\,12744---the first AR of the new solar activity Cycle\,25.
The goal of this work is to reveal the most characteristic features of the oscillation processes in a facula at an early stage of the activity cycle and to establish which frequencies dominate in the transition region and corona at different phases of the active region evolution.

\section{Instruments and Data}

We used data from the the \textit{Solar Dynamics Observatory} \citep[SDO,][]{2012SoPh..275....3P}.
The \textit{Atmospheric Imaging Assembly} (AIA) onboard SDO obtains full-disk images of the Sun in a number of ultraviolet spectral channels with a 0.6$''$ spatial resolution and 12 to 24\,s time cadence.
For the analysis, we chose the 1600\,\AA, 304\,\AA, and 171\,\AA\ channels, which represent the chromosphere, transition region, and corona, respectively.
We also used the \textit{Helioseismic and Magnetic Imager} (HMI) measurements of the line-of-sight magnetic field.

Ground-based observations at the \textit{Horizontal Solar Telescope} of the Sayan Solar Observatory complement the SDO data.
The working aperture diameter of the main mirror is 80\,cm; the Earth's atmosphere limits the spatial resolution to 1--1.5\,arcsec, and the spectral resolution varies between 4 and 8 m\AA\ depending on the wavelength.
A detailed description of the telescope can be found in \citet{Kobanov01, Kobanov13}.
The lengths of the series varied between 135 and 210\,min, and the cadence was 3.2\,s.
In the observations, we used three spectral lines: the chromospheric He\,\textsc{i} 10830\,\AA\ and H$\alpha$ lines and the photospheric Si\,\textsc{i} 10827\,\AA.

\section{Morphology and Evolution  of the Active Region}

The first AR of solar activity cycle 25 emerged at the disk on 6 July 2019 at the southern hemisphere close to the eastern limb (S27E55).
During the first six hours after the first appearance of the magnetic flux, it rapidly developed into a quite big facula (approximately 50$''$ in diameter based on the images in the 1600\,\AA\ channel).
After that, pores appeared in the AR.
The new AR was given number NOAA\,12744.

On 8 July, the facula started to expand.
It separated into head and tail parts of negative and positive magnetic polarities, correspondingly.
On this day, the pores disappeared, and the facula started to divide into several cores.

The 171\,\AA\ channel images showed a blurred picture of forming coronal loops.
The region continued to grow on 10 July 2019 in the 1600\,\AA\ and 304\,\AA\ images, while a clear structure of closed coronal loops formed in the 171\,\AA\ channel.

Starting from 11 July, the loop structures in the 171\,\AA\ images grew more blurred and elongated.
During the next two days, the loop structure continued to disintegrate, and by the end of this period (01:00 on 14 July), they disappeared completely.

\begin{figure*}
\plotone{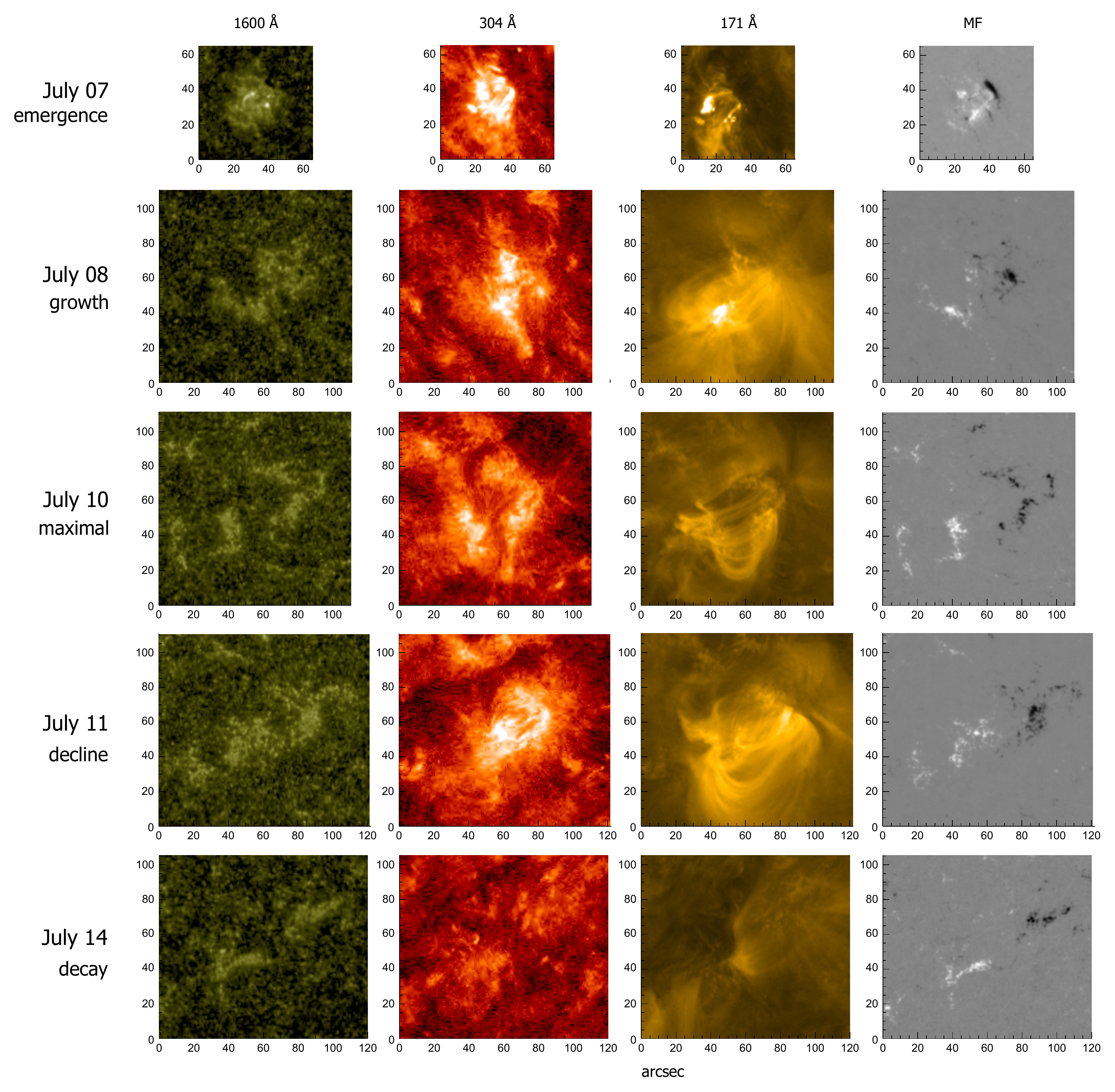}
\caption{Snapshots of the facula in the three AIA channels and HMI magnetograms for the five phases of the active region evolution.\label{fig:1}}
\end{figure*}

On the eighth day after the facula appeared, it decayed: it almost merged into the surrounding chromospheric network in the 1600\,\AA\ channel.

It must be noted that on 14 July at 01:00 UT after the decay, a new magnetic reconnection occurred followed by a local brightening in the coronal lines and in the upper chromosphere channels.
A new loop structure developed above the facula rapidly within an hour.
The newly activated region lived for two more days, during which the decay process repeated concluded by a complete disappearance of the active region.
By this time the region neared the western limb.

We divided the evolution of the facula into five phases based on the SDO/AIA images.
The first phase (\textbf{emergence}) is characterized by the appearance of small brightenings at an area of increased magnetic field concentration.
The second phase (\textbf{growth}) is distinguished by a rapid expansion of the bright elements.
At the third phase (\textbf{maximal development phase}), a stable coronal loop structure is observed in the 171\,\AA\ channel; the surface area and brightness of the facula are at their maxima.
The fourth phase (\textbf{decline}) is characterized by a decrease in the brightness and the number of the coronal loops.
At the fifth phase (\textbf{decay}), the coronal loop structure completely disappears.

We carried out the observations at the ground-based telescope on July 7th, 8th, 10th, and 11th, which corresponded to the emergence, growth, maximal development, and decline phases of the facula evolution.
We lack ground-based observation for the fifth phase---the decay.

\section{Results}
\subsection{Oscillation dynamics in the LOS velocity and intensity}

\begin{figure*}
\plotone{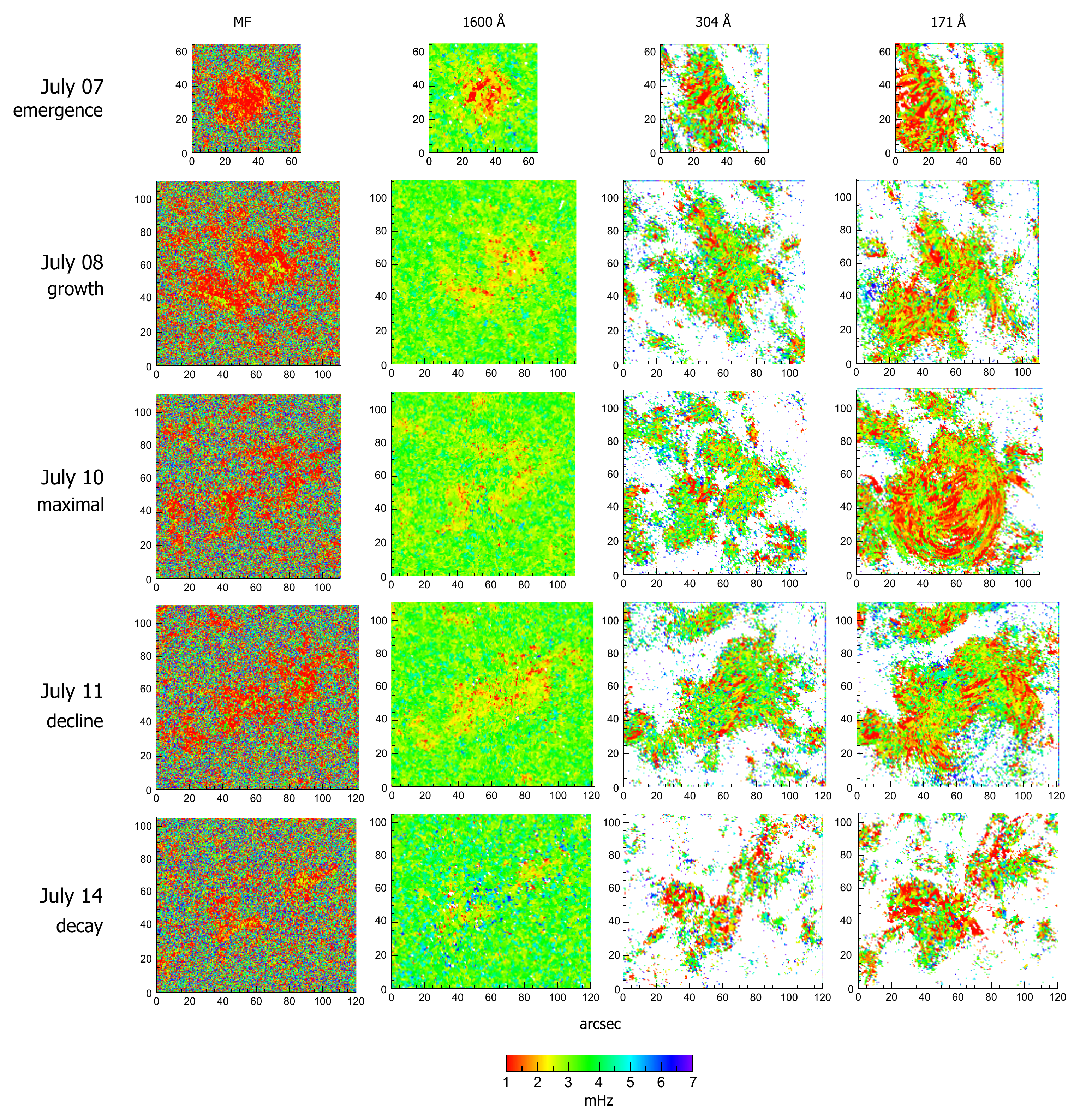}
\caption{Maps of the dominant frequencies in the faculae in the HMI LOS magnetic field signals and in the tree studied AIA channels at the five phases of its evolution.\label{fig:2}}
\end{figure*}

The AIA telescope onboard SDO provides two-dimension images in UV and EUV channels, which we used to construct spatial distributions of the dominant oscillation frequencies (Figure~\ref{fig:2}).
These distributions show which frequencies dominated at each point of the faculae during each of the five phases of the facula development.
We based the frequency distributions on the Fourier spectra.
The dominant frequency $\nu_d$ at each spatial point was determined as the frequency, for which the integrated oscillation power in the $\nu \pm$0.5\,mHz range shows the maximal value.
To clear the images of the noise, we filled with white the spatial points, whose oscillation power did not reach the 3$\sigma^2$ level.

For the four out of five facula evolution phases that we specified, we have ground-based spectral observations of the facula in the Si\,\textsc{i} 10827\,\AA\ (upper photosphere), He\,\textsc{i} 10830\,\AA\ and H$\alpha$ (chromosphere) lines.

\begin{figure*}
\plotone{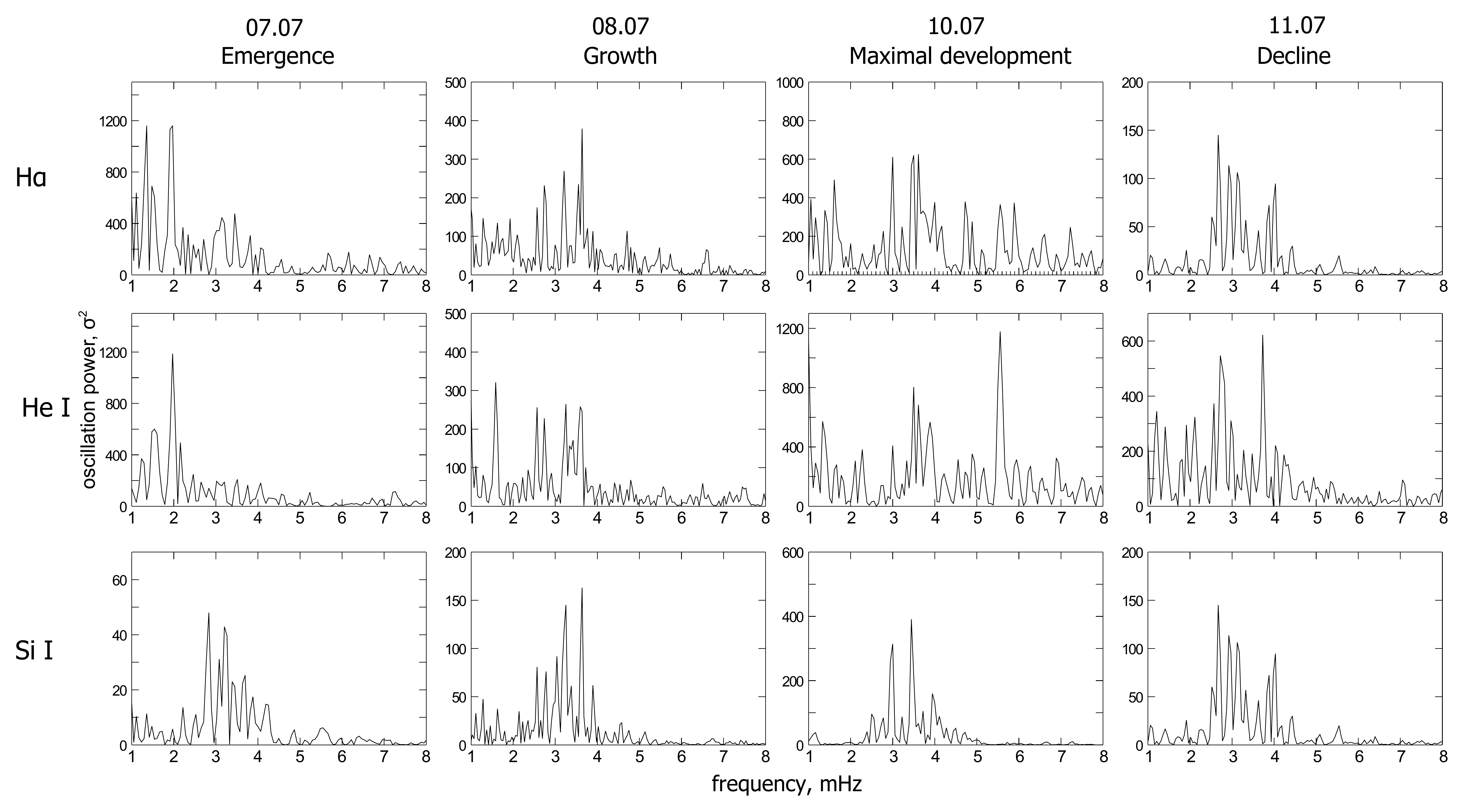}
\caption{Line-of-sight velocity oscillations in the central part of the facula based on the ground-based data.\label{fig:3}}
\end{figure*}

We analysed the spectra in the 1--8\,mHz range.
The restriction at the lower end was imposed by the fast evolution rate of the active region, which may lead to significant changes in its structure during the time series. 
The frequencies over 8\,mHz showed insignificant power.

Numerous earlier works have noted the dominance of the five-minute oscillations in faculae photosphere and chromosphere \citep{1974SoPh...39...31D,1974SoPh...39...79T, 1990SoPh..127..289B, 2005ApJ...624L..61D}.
In our case, however, the emergence phase showed the oscillation power concentration in the lower frequency range (1--2\,mHz) in the central part of the facula.
This clearly shows in the oscillation spatial distribution in the HMI magnetic field signals and in the 1600\,\AA\ (Figure~\ref{fig:2}).
The lower frequencies also dominate in the spectra of our ground-based LOS-velocity observations in the H$\alpha$ and He\,\textsc{i} 10830\,\AA\ lines over the central part of the facula at its first evolution phase (Figure~\ref{fig:3}).
Concentrations of elements with the lower-frequency oscillations are seen in the 304\,\AA\ and 171\,\AA\ as well.

Appearance of the lower frequencies in the chromosphere above faculae are mainly attributed to the decrease in the cut-off frequency in the areas of the inclined magnetic field \citep{1973SoPh...30...47M, 1977A&A....55..239B}.
However, \citet{2008ApJ...676L..85K, 2013JPhCS.440a2048K, 2009ApJ...692.1211C} indicate that the cut-off frequency may decrease even in the areas of the vertical magnetic field due to changes in the radiative relaxation time \citep{1983SoPh...87...77R}.
We believe that both effects could be working in facula atmospheres.

\begin{figure}
\plotone{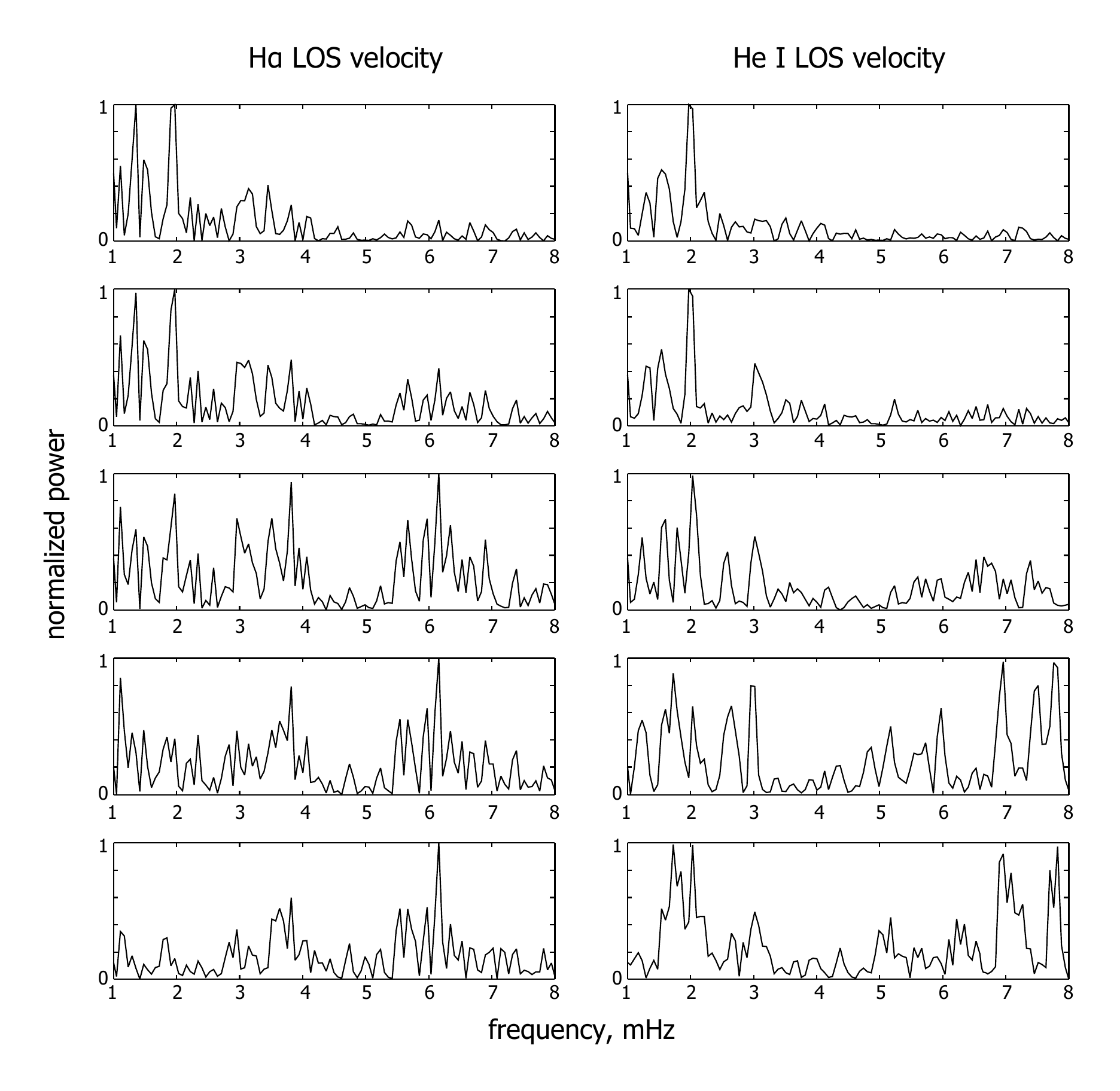}
\caption{Line-of-sight velocity oscillations in the chromosphere in the central part of the facula at its emergence phase and at points farther away from it (from top to bottom).\label{fig:4}}
\end{figure}

In the 1600\,\AA\ dominant frequency distribution of the emergence phase, a ring of higher frequency (up to 8\,mHz) interspersings  surround the core of the facula.
We observe the same pattern in the ground-based spectra of the chromospheric lines: at a distance of 10$''$ to 15$''$ from the central part of the facula, strong peaks manifest in the 6--8\,mHz range (Figure~\ref{fig:4}).

One should note that in this case we see a pattern opposite to that observed in sunspots, where high frequencies reside in the umbra's centre, and the dominant frequency gradually decreases closer to the outer edges of a sunspot; the lowest frequencies are found outside of the penumbra \citep{2000SoPh..196..129K, 2016SoPh..291.3339K}.
In sunspots, such a frequency distribution is attributed to the geometrical shape of the magnetic field---vertical at the centre and more and more inclined to the edges---and to the cut-off frequency related to it.

In the facula, however, the shift to the higher frequencies lacks the gradual transition through the middle-range frequencies.
The spectra show only two frequency domains with strong peaks: the lower frequencies around 2\,mHz and higher frequencies around 7\,mHz.
In the areas out of the facula's core, the high-frequency peaks grow, and the power in the central part of the spectra (3.5--5.5\,mHz) remains quite low in all the areas of the facula (Figure~\ref{fig:4}).
To answer the question whether such behavior is typical of the early stages of a solar cycle or it characterizes a birth of an active region at any stage of a cycle will require numerous observations of the oscillation dynamics in facular active regions throughout a whole cycle.

The magnetic field topology dramatically influences the distribution of the dominant frequencies in the facular photosphere and chromosphere.
\citet{2013A&A...559A.107K} studied the oscillation period dependence on the field strength in individual parts of a facula.
They concluded that the dominant oscillation period in the photosphere and low chromosphere increases in the areas with the increased strength of the underlying magnetic field.
\citet{2016SoPh..291.3329C} found that the power of the high-frequency oscillations increased in at the locations of strong magnetic field.
The mismatch of these results could be explained by the fact that \citet{2013A&A...559A.107K} used the full-vector data, while \citet{2016SoPh..291.3329C} used the LOS component.
The lower frequencies are usually associated with inclined field.
Besides, the observations in the two studies were carried out in different sets of spectral lines and with different series time lengths.
We should also note that the facula observed in the former study was at the last stage of its evolution: it had disintegrated; its surface area had decreased, and the magnetic field strength had weakened compared to its maximal development phase three days before.

Starting from the growth phase, the main oscillation power in the LOS velocity signals in the chromosphere shifted towards the five-minute range.
A similar shift is observed in the 304\,\AA\ and 171\,\AA\ channels (Figure~\ref{fig:2}), where in the dominate frequency distributions, the five-minute oscillations occupy extensive areas in the facula.

At the maximal development phase, the oscillation power in the 5--6\,mHz range increases in the lower levels.
This is especially noticeable in the LOS velocity signals of the He\,\textsc{i} line (Figure~\ref{fig:3}).
In the maximal phase, the coronal 171\,\AA\ line distribution in large part is occupied with the low-frequency oscillations (1--2\,mHz) in, which reproduces the loop structures that had formed by this time.
This suggests that in the 171\,\AA\ line we see the most horizontal apexes of the loops, and that the low-frequency oscillations prefer such structures.
Note that the spatial distribution of the low frequencies represents a picture averaged over the time series, while the images in Figure~\ref{fig:1} are one-moment snapshots, thus one should not expect a complete visual match.

At the last two evolution phases, the frequency distribution in the lower atmosphere levels is close to that in the elements of the undisturbed chromosphere network \citep{2005A&A...441.1191G, 2013SoPh..282...67G}.
The loop structure in the corona disappears, and the lower frequency areas become fragmented with higher concentration in the positive magnetic polarity area.
At the decay phase, the low frequency oscillations largely fade at the chromosphere level, and the higher frequencies of 5--7\,mHz evolve around the areas of the magnetic field concentrations.

In the photospheric Si\,\textsc{i} line, oscillations with 3--4\,mHz frequencies (five-minute oscillations) dominate at all the phases of the facula development.

In the spectra of the ground-based intensity signals, the main peaks are distributed in a wider frequency range.
In the photosphere, the oscillation power concentrates around the 3--4\,mHz, and in the chromosphere, the main peaks mostly seat in the 1--3\,mHz range.
In addition, the intensity spectra of the two chromospheric lines often differ significantly, as opposed to the velocity signals, where the oscillations of the He\,\textsc{i} and H$\alpha$ lines behave similarly.

\subsection{Oscillations in the LOS magnetic field signals}

The reality of the observed oscillations in the magnetic field signals is often doubted \citep{1998ApJ...497..464L, 1999ASSL..243..337R}.
Observations of oscillations in the magnetic field strength signals in faculae are rare.
Nevertheless, results of such observations in the 1--7\,mHz range are given in \citep{1995itsa.conf..387M, 2007SoPh..246..273K, 2016SoPh..291.3329C}. \citet{2016Ge&Ae..56.1052S, 2019Ap&SS.364...29S, 2019A&A...627A..10R} reported on longer periods of tens and even hundred minutes.
In our study, the changes in the magnetic field topology caused by the fast evolution of the active region restricted us to the 1--8\,mHz range.

The main part of the facula area in the distribution of the magnetic field oscillations is occupied with the low frequencies at the early stages of the facula development. The locations of the low frequencies correspond to the areas of the moderate magnetic field strength within the facula. By the time of the decay, the low and mid-range frequencies locate only close to the magnetic field concentration knots, and all the other areas between them are occupied with the multi-colored background, which corresponds to the signals of low oscillation power. A similar spectral composition fills the areas of the quiet Sun around the facula.

Since we analyse the line-of-sight component of the magnetic field strength, we should note that the angle at which the facula was observed changed in a quite wide range: the facula moved along the thirtieth parallel from meridian 50$^\circ$ East to the central meridian, and then to meridian 40$^\circ$ West.
Thus, the angle between the line of sight and the surface normal varied between 30$^\circ$ and 65$^\circ$. This could have influenced the observed shape of the magnetic field frequency distribution.

\subsection{Coronal loop oscillations}
In the spatial distributions of 171\,\AA\ channel dominant frequencies, the most prominent features are the structures of low frequencies that outline the coronal loops.
This stands out the most at the maximal phase of the facula evolution.
The presence of the low frequencies in the coronal loop spectra has been noted before \citep{2009A&A...503L..25W, 2002SoPh..206...99A, 2009ApJ...698..397V, 2018ApJ...854L...5D}. \citet{2015SoPh..290..363K, 2014ARep...58..272K} studied spatial distributions of oscillations above faculae; their analysis showed that the fan structures are clearly reproduced at the frequencies of 1--1.5\,mHz.

\citet{2011A&A...533A.116Y} found frequencies between 0.2 and 0.6\,mHz in the corona, as well as in the chromosphere above a sunspot, which indicates a long-period connectivity between these two layers.
They suggested that these waves might be incited in the sunspot magnetic flux tubes by the \textit{g}-modes.
Restricted by the length of the time-series, we observed shorter-period observations.
However one cannot exclude the same excitation mechanism, although in our case it is even more difficult to estimate the viability of this scenario, since the sub-photosphere magnetic field structure in faculae has been much less studied compared to sunspots.

Coronal loop spectra could be influenced, apart from the intrinsic intensity variations of the loop material, by the apparent changes in brightness caused by the transverse loop movements.
Such movements are present in the most active regions \citep{2015A&A...583A.136A}.
Even a sub-pixel displacement of a loop  changes the brightness of each separate pixel.
\citet{2016A&A...585A.137G} analysed 58 kink oscillation events and found that most often the 5-minute period oscillations occurred in them; they found the 8--12-minute oscillations as well.

Besides, the observed low frequencies could be related to the short lifetime of the loops.
To exclude this factor from the analysis, we restricted the analyzed time series to the periods when the loops existed stably at one location.
In the following example, we used a contrast coronal loop that was lit up during 32 minutes (02:13 to 02:45) at the maximal development phase.

\begin{figure}
\plotone{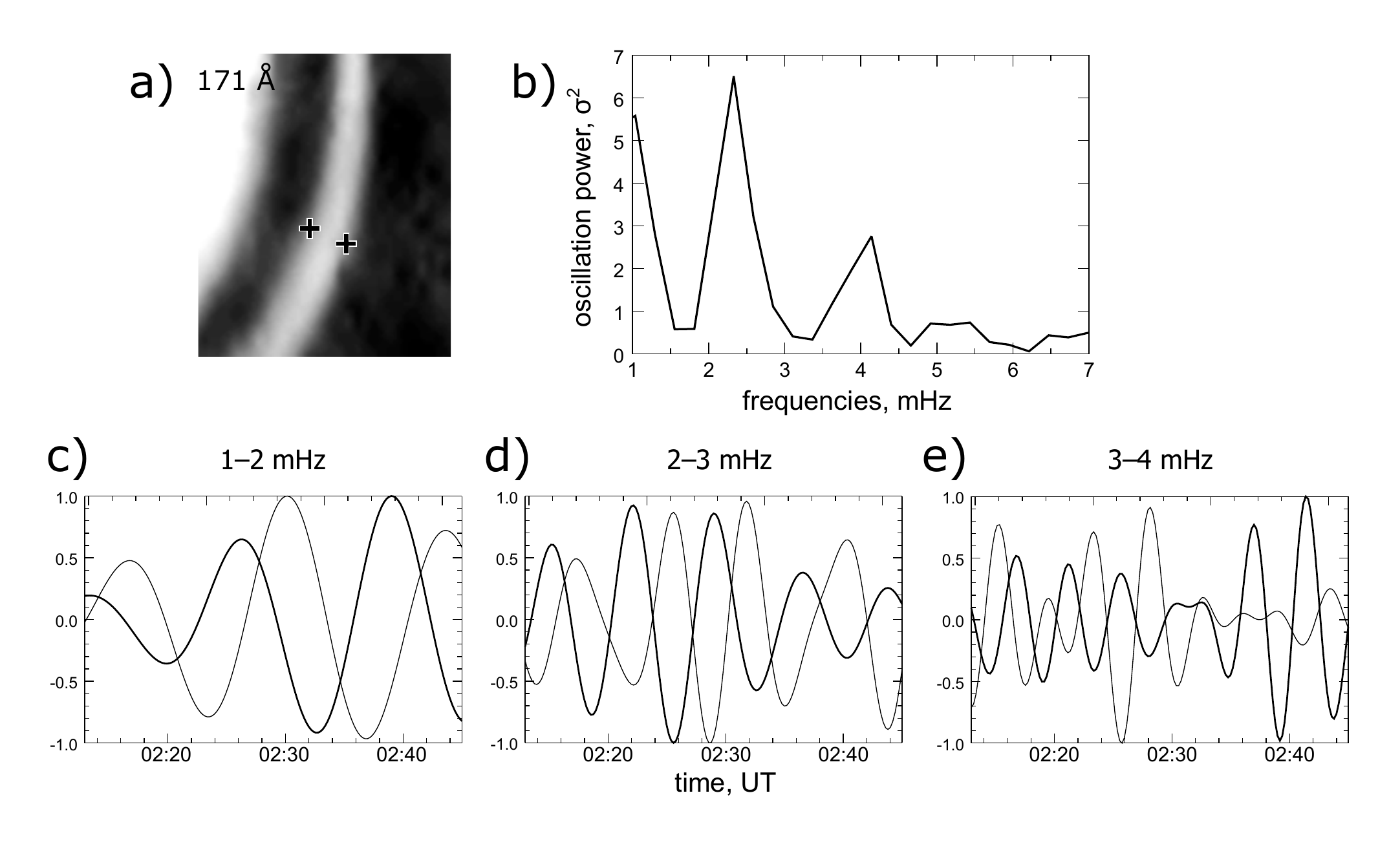}
\caption{a) the 171\,\AA\ loop image with the reference points on the sides of the loop marked with crosses; b) oscillation power spectrum of the 171\,\AA\ signal from a reference point; c-e) signals from the points on the sides of the loop wavelet-filtered in three frequency ranges.\label{fig:5}}
\end{figure}

To distinguish the oscillations related to the loop displacements, we analysed the signals from the points located on the sides of the loop central axis (Figure~\ref{fig:5}).
In such signals, the oscillations caused by the loop movements should be in anti-phase, while the oscillations of the loop brightness should be in phase.

The power spectra are dominated by the low frequencies accompanied by weaker peaks from 2 to 5\,mHz.
We used narrow-band wavelet filtration in order to separate different frequency oscilations.
The oscillations of the 2--3\,mHz and 3--4\,mHz ranges have the opposite phases.
This implies that they result from the transverse displacements of the loop.
The phase difference in the 1--2\,mHz band differs from 180$^\circ$, which indicates an input from other oscillations sources.

\section{Conclusion}

For the first time, we carried out a multi-wave research of oscillation dynamics in a solar facula from its birth to decay.
At the emergence phase of the facula evolution, the low-frequency (1--2\,mHz) oscillations concentrated in the centre of the facula in the photospheric and chromospheric signals of the intensity, LOS velocity, and LOS magnetic field.
This is the most apparent in the spatial dominant frequency distribution in the chromosphere 1600\,\AA\ channel.
The high-frequency (5--7\,mHz) oscillations group at the edges of the emerging active region.
We suggest that the occurrence of the low frequencies could be considered as a precursor for the coronal loop structure development in a growing facula.

Later, at the more developed phases of the facula evolution, 5-minute oscillations dominate the photosphere and chromosphere of the facula.

The spatial distribution of the dominant frequencies in the corona (171\,\AA) is closely related to the coronal loop dynamics in the facula.
During the maximal development of the coronal loop system these distributions are dominated by the low frequencies.
At these phases, the low frequency distribution closely resembles the coronal loops.
In the transition region frequency distributions, the low-frequency areas look more patchy: they concentrate around the footpoints of the coronal loops.

At the last phase of the evolution, the low frequencies in the chromosphere almost disappear, and the high frequencies take space between the magnetic field concentration areas.

Our results confirm the version that the sources of the low frequency oscillations in the coronal loops are located in the lower layers of the solar atmosphere.

\acknowledgments

The research was supported by the Russian foundation for Basic Research under grant 20-32-70076 and Project No.\,II.16.3.2 of ISTP SB RAS. Spectral  data were recorded at the Angara Multiaccess Center facilities at ISTP SB RAS. We acknowledge the NASA/SDO science teams for providing the data.

\end{document}